\documentclass[aps,prd,eqsecnum,amssymb,amsmath,showpacs, a4paper,superscriptaddress]{revtex4-2}

\usepackage{graphicx}
\usepackage{amsfonts}
\usepackage{amsmath}
\usepackage{color}
\usepackage{hyperref}
\usepackage{units}

\usepackage{dcolumn}
\usepackage{bm}
\usepackage{float}
\usepackage{natbib}
\usepackage{cleveref}
\usepackage[normalem]{ulem}
\usepackage{appendix}
\usepackage{mathtools}
\usepackage{esvect}
\usepackage[colorinlistoftodos, size=tiny, bordercolor=white]{todonotes}
\usepackage{comment}

\graphicspath{ {images/} }
\usepackage{mathrsfs}
\usepackage{amssymb}
\usepackage{dsfont}
\usepackage{enumitem}
\usepackage{textcomp, gensymb}

\usepackage{bm}

\usepackage{mathtools}
\usepackage{chngcntr}
\counterwithout{equation}{section}

\makeatletter
\let\cat@comma@active\@empty
\makeatother

\newcommand{\be}{\begin{equation}}
\newcommand{\ee}{\end{equation}}
\newcommand{\beal}{\begin{aligned}}
\newcommand{\eeal}{\end{aligned}}

\usepackage{xr}
\definecolor{rRGB}{RGB}{30, 144, 255}

\begin{document}
\setcounter{secnumdepth}{2}

\title{Astrophysical viability of extremal black holes}

\author{Chiara Coviello }
\affiliation{Department of Physics, King’s College London, University of London, Strand, London, WC2R 2LS, UK}
\email{chiara.coviello@kcl.ac.uk}
\author{Ruth Gregory}
\affiliation{Department of Physics, King’s College London, University of London, Strand, London, WC2R 2LS, UK}
\affiliation{Perimeter Institute, 31 Caroline Street North, Waterloo, ON, N2L 2Y5, Canada}
\email{ruth.gregory@kcl.ac.uk}

\date{\today}
\begin{abstract}
\noindent
We take a fresh look at the viability of physically realistic extremal black holes 
within our (non-supersymmetric) low energy physics. 
By incorporating prefactors and volume effects, we show that Schwinger discharge in charge neutral environments is far more efficient than commonly assumed. Using ionisation estimates for neutral hydrogen, we obtain a new and robust lower bound on the mass of an extremal electrically charged black hole, exceeding $10^{14} M_\odot$.
For magnetic black holes, we compute the Lee--Nair--Weinberg instability and revisit early universe pair creation rates, including singular instantons that substantially enhance production, to demonstrate that the extreme charges required for stability are cosmologically implausible.
Finally, we speculate that an extremal Kerr black 
hole could shed angular momentum via superradiant scattering from the stochastic gravitational wave background.
Taken together, our results provide a unified picture that extremal black holes are unlikely to persist in our universe.

\end{abstract}
\maketitle

\section{Introduction}

Extremal black holes remain among the most perplexing solutions in classical general relativity. 
They often sit at the boundary between naked singularities and well-behaved spacetimes, 
challenge our understanding of black hole thermodynamics, and yet we do not even know if they 
exist beyond our mathematical expressions.
In these objects, inner and outer horizons coincide, and they reach the maximum charge/angular 
momentum allowed for a black hole. 
Their vanishing Hawking temperature suggests a connection to the ill-understood third law of 
black hole thermodynamics, whereas their finite area raises puzzles with the conventional
interpretation of entropy as a counting of microstates.
It is precisely their unique thermodynamic character, as well as their threshold nature in terms of
cosmic censorship that makes them so compelling a study.

In a recent essay, Dafermos \cite{Dafermos:2025int} reflects on these aspects of extremality,
both discussing the historical development of our understanding as well as how and why that
understanding was incomplete.
The third law of black hole thermodynamics, as conjectured in \cite{1973threelaws}, simply
asserted that it should be impossible to reduce the surface gravity of a black hole to zero in 
a finite sequence of operations $-$ such a law would not only keep the thermodynamical 
analogy for black holes intact, but also would ensure that one could not, in a finite set of steps,
push a black hole beyond extremality, rendering it a naked singularity.
In 1986, Israel \cite{Israel3law}, provided a definitive formulation of this third law, then presented
a proof that there was no {\it continuous} process in which the black hole could be rendered 
extremal while maintaining the weak energy condition. However, the results of Sorce and Wald
\cite{2017GedankenWald2}, while focused on proving that no gedanken experiment can succeed in 
overcharging or overspinning a Kerr--Newman black hole, suggest that an extremal black hole can be 
reached from a near-extremal one via carefully tuned perturbations, assuming only the null energy 
condition. Indeed recently, Kehle and Unger \cite{Kehle:2025} have shown how to form 
an extremal Reissner--Nordstr\"om black hole through gravitational collapse in 
Einstein--Maxwell--charged scalar field theory $-$ we refer the reader to \cite{Dafermos:2025int} for
a comprehensive yet concise critical evaluation of these results and how Israel's original proof
is evaded. 

Beyond their thermodynamic peculiarities, extremal black holes play a key role in 
fundamental high-energy physics. They often emerge as supersymmetric gravitational solutions in 
string theory \cite{Gibbons:supersymmetry,Horowitz:strings}. 
Indeed, a recent tantalising result of Brown et al.\ \cite{BrownEvaporationCHB}
explores the partition function of 
near extremal black holes and, via an exact computation in JT gravity, argue
that as one approaches extremality, quantum gravitational effects strongly modify
the partition function, suppressing the entropy. This would appear to resolve the 
puzzle of an extremal black hole having finite area
yet being at zero temperature that was debated some years ago in the literature
(see e.g.\ \cite{Hawking:1994ii,Ghosh:1996gp}).
The notion that an extremal, or near extremal, black hole should receive 
strong quantum corrections then raises the question of whether these modifications
could be observable, and indeed could provide a clear signal of these corrections
\cite{Emparan:2025sao,Emparan:2025qqf} $-$ a tantalising prospect indeed!

The discussion of Dafermos (and indeed many explorations of over-charging or spinning 
black holes such as \cite{WALD1974Gedanken1, Hubeny, 2017GedankenWald2}) all
treat the black hole in splendid isolation $-$ the spacetime is determined by classical 
general relativity with test particles or Maxwell fields. In our universe however, black holes are 
not isolated systems, therefore we critically re-examine the question:
can extremal black holes exist in our universe? 
This is a crucial question, as the answer determines whether such geometries can serve as 
a potential testing ground for quantum gravity and black hole thermodynamics, or whether extremality 
is inherently unreachable, with black holes naturally evolving away from it, rendering these 
spacetimes inaccessible and ultimately unobservable.

The possibility of long-lived electrically charged black holes is often dismissed 
on the grounds that the Coulomb force vastly exceeds gravity, implying that charged 
particles in any surrounding plasma would rapidly neutralize the black hole \cite{1975ARA&A..13..381E}. 
Here we consider the viability of physically realistic charged extremal black holes
immersed in a \textit{neutral} environment (the QED vacuum and neutral hydrogen). 
We first revisit the Gibbons \cite{Gibbons:schwingerpairs} Schwinger discharge
process, re-evaluating the probability of discharge including 1-loop effects and a 
proper consideration of the probability of discharge. We then derive an even
stronger constraint from the ionisation of matter in the local environment.
In both cases, the impact of the prefactor and volume in which electric discharge 
occurs is extremely significant. 
We then evaluate the likelihood of extremal magnetic black holes, that might be expected to be
stable to any pair production decay, computing the details of the linear instability first argued
by Lee, Nair and Weinberg \cite{Lee:1991}. We also revisit arguments on black hole creation during
inflation, noting a broader range of instantons than previously considered.
We conclude with remarks on Kerr extremality.

\noindent{\it A remark on units}

Throughout our discussion, we will, in the main, use natural units as is standard convention
in GR literature, i.e.\ $\hbar=c=4\pi\epsilon_0=1$. We will also use the standard (as opposed to
reduced) definitions of Planck units, setting $G=1$. However when discussing the details of 
physical processes, where scales and numerical estimates are relevant, we restore these 
constants in the presentation of results, using SI units.

\section{Electrically charged extremal black holes}
\label{sec:electric}

We first discuss the feasibility of an extremal electrically charged black hole.
A classic result by Gibbons \cite{Gibbons:schwingerpairs} calculated how the electrostatic field of an 
electrically charged black hole can induce Schwinger pair production, leading to charge depletion.
The essence of this effect is that the energy of an electric field 
is proportional to $|{\bf E}|^2$, thus the stronger the electric field, the greater the drop in energy 
if the field discharges an amount $\delta {\bf E}$.
Meanwhile, to create an electron/positron pair, we require (at minimum) a rest
mass energy of $2m_ec^2$. An estimate of the drop in the electric field over a distance $d$ due
to the pair would have the dependence $\sim |{\bf E}_0| e d$, where we can estimate 
$d$ using the uncertainty relation $\hbar \sim \delta E \delta t \sim m_e c^2 d/c$. 
Thus we expect pair creation to start to discharge
the electric field at a magnitude around
\be
|{\bf E}_0| \sim {\cal O} \left ( \frac{m_e^2 c^3}{e \hbar} \right)
\ee
consistent with the Schwinger result for the decay rate per unit volume:
\be
\Gamma_{\text{Schwinger}} = \frac{(e|{\bf E}_0|)^2}{4\pi^3\hbar^2 c}
\sum_{n=1}^\infty \frac{1}{n^2} \text{exp} \left [ - \frac{n \pi m_e^2 c^3 }
{\hbar e |{\bf E}_0|} \right] 
\label{SchwingerGamma}
\ee

Putting this into the context of a charged black hole, the electric field scales 
approximately as $Q/r_h^2$, the charge of the black hole divided by the square of the horizon radius.  
For an extremal black hole, $r_h \sim Q \sim M$, hence $|{\bf E}| \propto 1/M$. 
Thus, smaller black holes will have stronger local electric fields and will be more likely to discharge 
leading to a lower bound on the allowed mass of an extremal black hole.
This is the intuition behind the argument of Gibbons, however the results / estimates above
were all obtained without consideration of gravity, or the geometry of the black hole $-$ indeed 
the original motivation of Gibbons was to understand particle production processes in strong
gravitational fields.

In order to determine the discharge of an electrically charged black hole, Gibbons followed
a similar argument to that of Hawking \cite{1975Hawking:paircreation} $-$ computing the Bogoliubov coefficients between in
and out states modelled with a charged scalar quantum field for simplicity. By considering
small and large black hole mass limits, he confirmed the validity of the (flat space) Schwinger
result \eqref{SchwingerGamma} for large black hole masses -- as might be anticipated from the small local 
spacetime curvature \footnote{{The flat space Schwinger formula is expected to hold when 
the spacetime curvature is negligible over the quantum tunnelling length scale, i.e.
$\lambda_e \ll \mathcal{R}_{\rm curv}$, where $\lambda_e = 2\pi\hbar/m_e c$ is the 
electron Compton wavelength. At all mass scales relevant to our analysis, the curvature 
scale exceeds this by many orders of magnitude; e.g.\ for the limiting mass we derive, $M=10^7\, M_\odot$, this ratio is $\lambda_e /\mathcal{R}_{\rm curv}\approx 10^{-22}$}}.
We therefore revisit the Schwinger discussion for large extremal black holes, noting the 
importance of the prefactor, which has been overlooked in the literature for this purpose to the best
of our knowledge. 

Recall that the Schwinger 1-loop computation gives a probability per
unit volume per unit time. In order to decide how probable it is therefore for a given 
extremal black hole to discharge one unit of charge, we must take this probability density
and multiply by the effective 4-volume in which the discharge can take place.
To get a first estimate of the likelihood of discharge for an extremal black hole, we
evaluate the Schwinger amplitude at the event horizon of the black hole, 
and multiply by the representative 4-volume $r_h^4/c$.
This factor is chosen to approximate the local 4-volume of the near
horizon region, taking a time-interval linked to the black hole of its light-crossing time
\footnote{The volume is estimated using local Kruskals, these are 
distinct from the usual expressions \cite{Kruskal:1959vx}, 
and were derived by Carter \cite{Carter:1966zza},
but it is straightforward to transform to
this system and confirm that $r_h^4/c$ captures the correct behaviour of this volume.}.
The idea behind this factor is that discharge will occur primarily in the near-horizon r\'egime
and that a time interval linked to the black hole removes any ambiguity of choosing a representative
time of an observer. Note that this time interval is typically short on astrophysical scales, 
making the bound that we get very conservative. For example, for an extremal black hole with mass $M=10^7\,M_\odot$ has a light-crossing time of order a minute.

Restoring units, the location of the horizon of an RN black hole is given by
\be
1 - \frac{2GM}{c^2 r} + \frac{GQ^2}{4\pi\varepsilon_0c^4r^2} =0 \;,
\ee
thus the extremal limit occurs when
\be
Q = \sqrt{4\pi \varepsilon_0 G} M \;\;\; ; \qquad
r_h = \frac{GM}{c^2} = \sqrt{\frac{G}{4\pi \varepsilon_0}}\, \frac{Q}{c^2}
\ee
Meanwhile, the electric field for a black hole with charge Q is 
\be
E_r = \frac{Q}{4\pi\varepsilon_0 r^2}
\ee
Thus, substituting in \eqref{SchwingerGamma} we obtain:
\be
\Gamma_{\text{Schwinger}} = \frac{(eQ)^2}{64\pi^5\varepsilon_0^2\hbar^2 c r^4}
\sum_{n=1}^\infty \frac{1}{n^2} \text{exp} \left [ - \frac{4n \pi^2 \varepsilon_0 m_e^2 c^3 r^2}
{\hbar e Q} \right] 
\label{eq:SCH(r)}
\ee
Multiplying by the volume factor, we obtain that the overall likelihood of the black hole discharging 
is of order:
\be
\beal
\mathcal{P} \sim \frac{r_h^4}{c}\, \Gamma_{\text{Schwinger}} 
&= \frac{(eQ)^2}{64\pi^5\varepsilon_0^2\hbar^2 c^2}
\sum_{n=1}^\infty \frac{1}{n^2} \text{exp} \left [ - \frac{n \pi m_e^2 G Q}
{\hbar e c} \right ]\\
&= \frac{\alpha^2}{4\pi^3} \frac{Q^2}{e^2} \sum_{n=1}^\infty \frac{1}{n^2} \text{exp} 
\left [ - \frac{4 n \pi^3 L_p^2}{\lambda_e^2} \frac{Q}{e} \right ]
\eeal
\label{Schwingerampl}
\ee
where $\alpha=e^2/4\pi\varepsilon_0 \hbar c$ is the fine structure constant, 
$L_p^2 = G \hbar/c^3$ is the Planck length, and $\lambda_e = 2\pi \hbar / m_e c$
is the Compton wavelength of the electron.

Expressed in this form, the dependence on the physical quantities is made manifest.
The hierarchy between the Planck length and the Compton
wavelength of the electron indicates the large strength of the electric field
required to damp the exponential factor (recall this is an extremal black hole so that
the strength of the electric field varies inversely to the black hole size). 
However the prefactor also plays an important role: the value of the 
fine structure constant gives a small numerical factor, $\alpha^2/4\pi^3 \sim 4
\times 10^{-7}$, but $Q/e$ is extremely large before the exponential factor 
starts to damp the pair creation. The parametric dependence therefore of the 
critical electric field is set by the Lambert function, or the inverse of $ze^z$, 
for the primary $n=1$ term in the Schwinger expression:
$\mathcal{P}\approx 1$ for
\be
\beal
\left ( - 2\pi^3 \frac{L_p^2}{\lambda_e^2} \frac{Q}{e} \right) \text{exp}
\left [  - 2\pi^3 \frac{L_p^2}{\lambda_e^2}  \frac{Q}{e} \right ] 
&= - \frac{4 \pi^4 \sqrt{\pi} L_p^2}{\alpha \lambda_e^2} \\
\Rightarrow \qquad \qquad
 - 2\pi^3 \frac{L_p^2}{\lambda_e^2} \frac{Q}{e} &= W_{-1} \left ( 
- \frac{4 \pi^4 \sqrt{\pi} L_p^2}{\alpha \lambda_e^2} \right )\\
\Rightarrow \qquad \qquad
\frac{Q}{e} &\approx 3.6 \times 10^{46}
\eeal
\label{eq:limiting-charge-schwinger}
\ee
This value corresponds to 
\begin{equation}
    M \approx 3.4 \times 10^7 M_\odot,
\end{equation}
or two orders of magnitude
higher than the threshold value obtained by considering the exponent dependence alone. 

Clearly this is an estimate, constructed to account for the main dimensional quantities 
influencing the decay, since it gives a substantially higher limit for stability against
discharge, a more careful discussion is in order. 
First consider the volume factor being used for the estimate;
rather than just multiplying the horizon rate 
by a representative factor, one should integrate the local rate outside the 
horizon in an appropriate measure. 
The extremal black hole in $\{t,r\}$ coordinates has an infinite
``throat'' in the radial direction that would appear to give a large enhancement,
however, the $t-$coordinate is correspondingly redshifted nullifying this effect.
To avoid issues with singular coordinate systems, we use ingoing Eddington--Finkelstein
coordinates, which are regular at the future horizon: 
\begin{equation}
    ds^2=
    -\left(1-\frac{r_h}{r}\right)^2 dv^2
    +2\,dv\,dr+r^2 d\Omega^2 ,
\end{equation}
where $v  = t + \int r^2/(r-r_h)^2 dr$ is the advanced time. Note that since
$\Delta v = \Delta t$, we should integrate out from the black hole horizon to an asymptotic
region, and over an interval $\Delta v = \Delta t$.

One might also worry that the flat space
Schwinger rate is not reliable due to corrections from this 
${\rm AdS}_2 \times {\rm S}^2$ approximate near horizon geometry.
A useful check is provided by the analysis of Lin and Shiu
\cite{Lin:2024jug}, who used a worldline
instanton approach to carefully follow the exponent and prefactor 
outside the horizon. From figure 2 in \cite{Lin:2024jug},
one sees that the exponent roughly doubles at a distance of half the horizon radius
from the horizon, which is broadly in keeping with the behaviour of the electric
field in the exponent. Similarly, in figure 4 they present the radial behaviour of the 
prefactor as one increases the charge:mass ratio, from which we see that the prefactor 
increases with increasing charge:mass ratio and drops inversely as one moves away 
from the horizon. While these results were primarily presented for particles whose 
charge:mass ratio was of order unity, they found that the prefactor 
scales as $\sim \sqrt{z^2-1}$, where $z$ is the charge:mass ratio. Here, we are
focussing on electron-positron pairs as they are the most likely to cause discharge,
thus we are well within the large charge:mass ratio regime: $z\sim10^{21}\gg1$.

The most significant departure of the Lin-Shiu result from \eqref{SchwingerGamma}
is a mild divergence of the prefactor $\propto 1/\sqrt{r-r_h}$ close to the 
horizon due to the soft modes present in the throat, however, on integrating this
near horizon result, a finite rate is obtained that differs from our estimated rate
by approximately a factor of 2, which does not significantly change our conclusions
(see discussion below).

To get the integrated Schwinger discharge likelihood, we define our variables in terms of horizon scales:
\be
S_0 = 4 \pi^3 \frac{L_p^2}{\lambda_e^2}\frac{Q}{e}\;, \qquad
A_0 = \frac{\alpha^2}{\pi^2} \left ( \frac {Q}{e} \right ) ^2
\label{exp-prefactor}
\ee
and let $\zeta=r/r_h$ be our radial integration variable so that the
Schwinger rate can be expressed as:
\be
\Gamma_{\rm Schwinger}(r)= \frac{c}{4\pi r_h^4} 
\frac{A_0}{\zeta^4} \text{exp} [-S_0 \zeta^2].
\ee
We then multiply by the volume measure in Eddington-Finkelstein coordinates, and integrate out to the asymptotic region for a time interval 
$\Delta v = \Delta t$ to obtain the discharge probability ${\cal P} = 
\text{Min}[1,{\cal P}_{\text{int}}]$, where
\be
\beal
\mathcal{P}_{\text{int}} &= 
\int \sqrt{|g|}\, \Gamma_{\rm Schwinger}(r) d^4x
= \frac{\Delta t}{r_h/c} A_0 \int_1^\infty
\frac{e^{-S_0 \zeta^2}}{\zeta^2}d\zeta \\
&=
\frac{\Delta t A_0}{r_h/c} \left ( e^{-S_0} -
\sqrt{\pi S_0} \, \text{Erfc}[\sqrt{S_0} ] \right )
\eeal
\label{intschwinger}
\ee
with $\rm Erfc$ the complementary error function. 
Choosing $\Delta t$ to be the light crossing time, and solving $\mathcal{P}_{\rm int}=1$ we get the limiting charge: 
\begin{equation}
    \frac{Q}{e}\approx 3.6 \times 10^{46},
\end{equation}
which is the same (to two significant figures) as the one found via our approximation. 

It is at first surprising that a more careful analysis with a proper
measure and integration should give the same as an extremely crude volume
estimate. However, the reason lies with the prefactor, and the mathematics of the 
expression. First of all, note that although $S_0$ grows linearly with 
$Q/e$, the gradient of $4\pi^3 L_p^2/\lambda_e^2 = 5.5 \times 10^{-45}$, thus $Q/e$ has to be
extremely large before $S_0$ becomes appreciable. At this point the prefactor has 
become large since it is proportional to ${\cal O}(10^{-7} (Q/e)^2)$. 
Thus, the exponential only starts to damp the prefactor at large $S_0$. 
Therefore, we can expand \eqref{intschwinger} at large $S_0$ to yield:
\be
\mathcal{P}_{\text{int}} = \frac{\Delta t A_0}{r_h/c} \frac{e^{-S_0} }{2S_0}
= \frac{A_0}{2S_0^2} S_0 e^{-S_0} \quad \text{for} \;\;\; \Delta t = r_h/c
\ee
and see that the expression for the black hole charge again involves the 
Lambert function 
\be
4 \pi^3 \frac{L_p^2}{\lambda_e^2}\frac{Q}{e} = - W_{-1} \left [ -
\frac{32 \pi^8 L_p^4}{\alpha^2 \lambda_e^4} \right ].
\label{eq:SchwLamb-int}
\ee
Defining $\epsilon = 2 S_0^2/A_0 = 32 \pi^8 L_p^4/\alpha^2 \lambda_e^4 \ll 1$, 
we can now understand why the bound on $Q/e$ is fairly insensitive to
moderate numerical scaling of the integrated discharge probability.
Using the expansion of the relevant branch of the Lambert function 
we see that 
\be
S_0 = - W_{-1}[-\epsilon] 
\simeq - \log \epsilon + \log[-\log\,\epsilon] + ...
\label{fullexpansion}
\ee
whereas in expression in \eqref{eq:limiting-charge-schwinger} we have
\be
\frac{S_0}{2} e^{-S_0/2} = \frac{4 \pi^{9/2} L_p^2}{\alpha \lambda_e^2} 
= \sqrt{\frac{\pi \epsilon}{2}} \;\;\; \Rightarrow \quad
\frac{S_0}{2} = - W_{-1} \left [-\sqrt{\frac{\pi \epsilon}{2}} \right  ]
\ee
giving
\be
\beal
S_0 &= - \left ( \log \epsilon + \log(\pi/2) \right ) +
\log \left [ - \left ( \log \epsilon + \log(\pi/2) \right ) \right ]\\
&= - \log \epsilon - \log(\pi/2) + \log[-\log\,\epsilon] - \frac{\log(\pi/2)}{\log\epsilon}.
\eeal
\label{roughexpansion}
\ee
Since $-\log\epsilon = {\cal O}(10^2)$, whereas $\log(\pi/2)\sim 0.45$,
it is easy to see that \eqref{roughexpansion} differs from \eqref{fullexpansion}
at a sub-leading order (around the $1\%$ level) and the critical black hole 
charge is dominated simply by the
leading order behaviour of $\epsilon$. Note, this also justifies the
use of the simpler flat space Schwinger amplitude as the Lin-Shiu corrections
simply give an order unity correction in the prefactor element that does
not impact on this Lambert computation.

The bound presented above has been computed for a time interval equal to the light-crossing time for the black hole. In cosmological terms, this is a very small number!
A better question might be {\it ``How long does an asymptotic observer have to wait for the black hole to discharge one unit and become non-extremal?''}
This can easily be determined from our expression \eqref{intschwinger} by 
re-arranging to solve for $\Delta t$:
\be
\Delta t = \frac{r_h}{c A_0}
\left ( e^{-S_0} -
\sqrt{\pi S_0} \, \text{Erfc}[\sqrt{S_0} ] \right )^{-1}.
\label{timescale}
\ee
\begin{figure}[ht]
    \centering
    \includegraphics[width=0.8\linewidth]{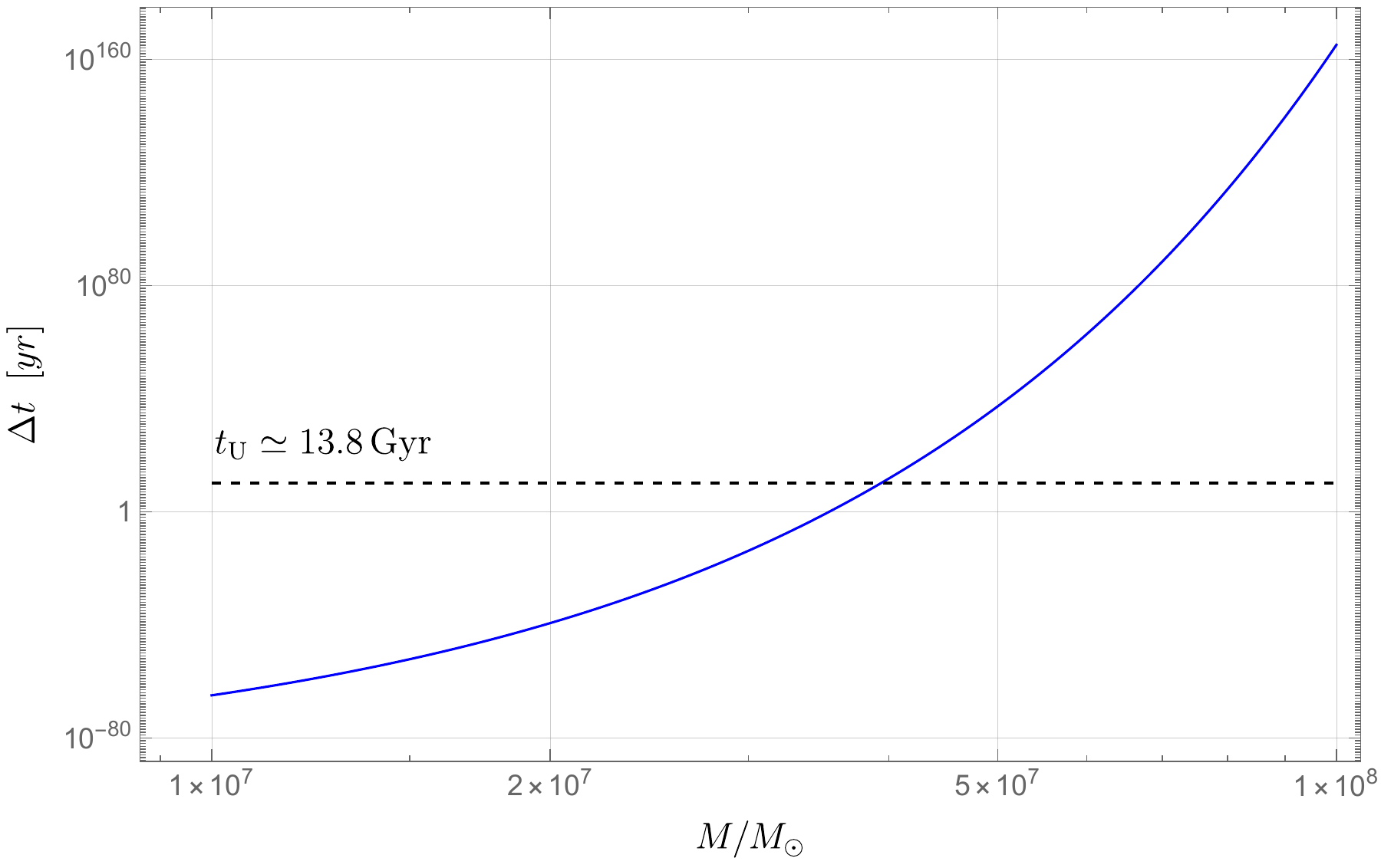}
    \caption{{ {\it Discharge timescale due to Schwinger pair production.} Plot of the timescale (in years) for Schwinger discharge of an extremal electrically charged black hole as a function of its mass. For each mass, the charge is fixed by extremality and $\Delta t$ is obtained from \eqref{timescale}. The dashed horizontal line indicates the age of the Universe, $t_U\simeq 13.8\,{\rm Gyr}$.
    }}
    \label{fig:timescaleSchwinger}
\end{figure}
Figure~\ref{fig:timescaleSchwinger} shows the dependence of this time interval on the 
mass of the black hole. It is striking how dramatically the timescale for decay 
increases for very small relative changes in the mass of the black hole, or conversely
how robust the mass limit is.
Previously, we had obtained a lower bound of $3.6\times 10^7 M_\odot$ on the mass for 
an extremal black hole to be stable against discharge on a timescale of order the 
light crossing time of the black hole. What figure~\ref{fig:timescaleSchwinger} shows 
is that even if we allow a cosmological timescale for decay, this limit is only 
increased to about $3.9\times 10^7 M_\odot$, and black holes
even a small amount above this mass are stabilised.
Thus, we may reasonably deduce that only the most massive supermassive black holes are 
sufficiently large to evade the Schwinger process. 

At the time of Gibbons’ paper (1975), the existence of black holes was under debate, at
least within the gravity theory community $-$ witness the famous wager between Stephen 
Hawking and Kip Thorne in that same year as to whether black holes existed in our universe! 
However, in general it was thought that black holes would mostly form from gravitational collapse of 
stars at the end of their lifetimes, so such a large limit for the discharge of a black hole 
effectively meant that extremal black holes were just not a possibility. Indeed, in
\cite{Carr:1974nx}, Carr and Hawking, while allowing for the possibility of 
\emph{primordial black holes} formed in the very early universe, also argued that
black holes would not accrete substantially, hence very large mass black holes were unlikely.

The modern perspective however is very different. Although extremely massive black holes
had been posited to sit at the centre of galaxies by astronomers for some years (see e.g.\
\cite{1971MNRAS.152..461L}), the community at large accepted this as fact only after
the observation of stars, in particular S2, 
orbiting the centre of the Milky Way provided the irrefutable evidence that clinched the 
case for \emph{supermassive black holes} (SMBH) \cite{Eckart:1996zz,Ghez:1998ph} (though it took
Hawking another 6 years to concede his bet with Thorne). Since then, the observational
evidence for black holes has become overwhelming, with the LIGO gravitational wave
data from black hole mergers \cite{LIGOScientific:2016aoc} and the Event Horizon Telescope 
imaging of the central SMBHs in our own galaxy and M87 
\cite{EHTsgrA,EHTM87} $-$ as well as exquisite imaging
of the stellar motion around our own black hole by the Gravity consortium \cite{Gillessen:2008qv}.

Since such large mass black holes are now believed to be commonplace, it is
(very) theoretically possible that there could be an extremal black hole that might
exceed the Schwinger bound and remain extremal, or close to (see the discussion above
and also \cite{BrownEvaporationCHB}).
However we now argue that the required electric field for protection from discharging
is in fact even \emph{lower} than the Schwinger bound, as black holes in our universe are not
in vacuum, but in an environment with dust and gas. Indeed, the very observations that 
confirm their existence rely on their interactions with their surroundings: their gravitational 
influence shapes the orbits of nearby stars, while their accretion disks $-$ composed primarily 
of hydrogen and helium, along with heavier elements in trace amounts $-$ emit radiation 
across the electromagnetic spectrum. 
Imaging of SMBHs captures photons emitted by this infalling material 
as it heats up and radiates. Therefore, even if the Schwinger pair production is inefficient, 
discharge can still proceed via ionisation of the surrounding matter, if that is not already ionised. Thus, 
although the electric field of an extremal SMBH is weaker than that of 
a smaller one, it remains sufficiently strong to ionise hydrogen $-$ the most abundant element 
in galactic nuclei and accretion environments.

For simplicity, we take hydrogen to be the dominant baryonic component of the local environment around the SMBH. The neutral hydrogen estimate below should be interpreted conditionally: if neutral atoms are present, the black hole electric field ionises them and contributes to discharge; if the gas is already ionised, the appropriate channel is instead direct plasma accretion, which is also expected to neutralise the black hole efficiently \cite{1975ARA&A..13..381E}. 

Here, we focus on the ionisation probability of neutral hydrogen.
The tunnelling ionisation rate, $p$, for an individual hydrogen atom in its ground state 
in a strong electric field $|{\bf E}_0|$ can be read off from \cite{LandauLifshitz} as: 
\be
\beal
p & =\frac{4 m_e^3 e^9}{(4\pi \epsilon_0)^5  \hbar^7|{\bf E}_0|}
\exp\left(-\frac{2 m_e^2 e^5}{3 (4\pi\epsilon_0)^3 \hbar^4|{\bf E}_0| }\right)\\
&= 32 \pi^3 \alpha^5 \frac{c L_p^2}{\lambda_e^3} \frac{Q}{e}
\exp\left(-\frac{8 \pi^2 \alpha^3}{3 } \frac{L_p^2}{\lambda_e^2} \frac{Q}{e} \right).
\eeal
\label{ionampl}
\ee 
where in the last step we have substituted the expression of the electric field at the horizon. Note the similarity to the Schwinger amplitude \eqref{Schwingerampl}, but this exponent
has the additional damping factor of $\alpha^3$. 

As with the pair creation estimate, note that this is the probability of ionisation
of a given atom per unit time. Therefore, as before, to properly estimate the likelihood of 
black hole discharge, we take into account the number of atoms in the vicinity of the black
hole, and a representative timescale, which we initially take to be
$r_h/c$ as before.
{We adopt the number density of hydrogen atoms to be around
$n \sim 10^6 cm^{-3}$ \cite{Smith_2013}, as a representative density for dense gas in galactic-centre environments, and consider the atoms to be spread in an $r_h^3$ volume. Observations and modelling of the galactic centre circumnuclear disk find densities ranging from roughly $10^5$ to $10^6\, \text{cm}^{-3}$ in dense molecular components, although lower density phases are also present. The dependence of the limiting charge (and so mass) of the black hole on $n$, as we will show, is however very weak.}
Thus, we multiply the individual rate of ionisation by the factor
\be
n \frac{r_h^4}{c} = n \frac{G^2 Q^4}{(4\pi\varepsilon_0)^2 c^5} = 
n \alpha^2 \frac{L_p^4}{c} \left ( \frac{Q}{e} \right)^4
= {\cal N} \frac{\alpha^2 L_p}{c} \left ( \frac{Q}{e} \right)^4
\ee
where ${\cal N} = n L_p^3$ is the number of atoms per Planck 
volume near the black hole. {In the fiducial case with $n=10^{-6}\,\rm cm^{-3}$, we get ${\cal N}\sim 4 \times 10^{-93}$.}

\begin{figure}[ht]
    \centering
    \includegraphics[width=0.8\linewidth]{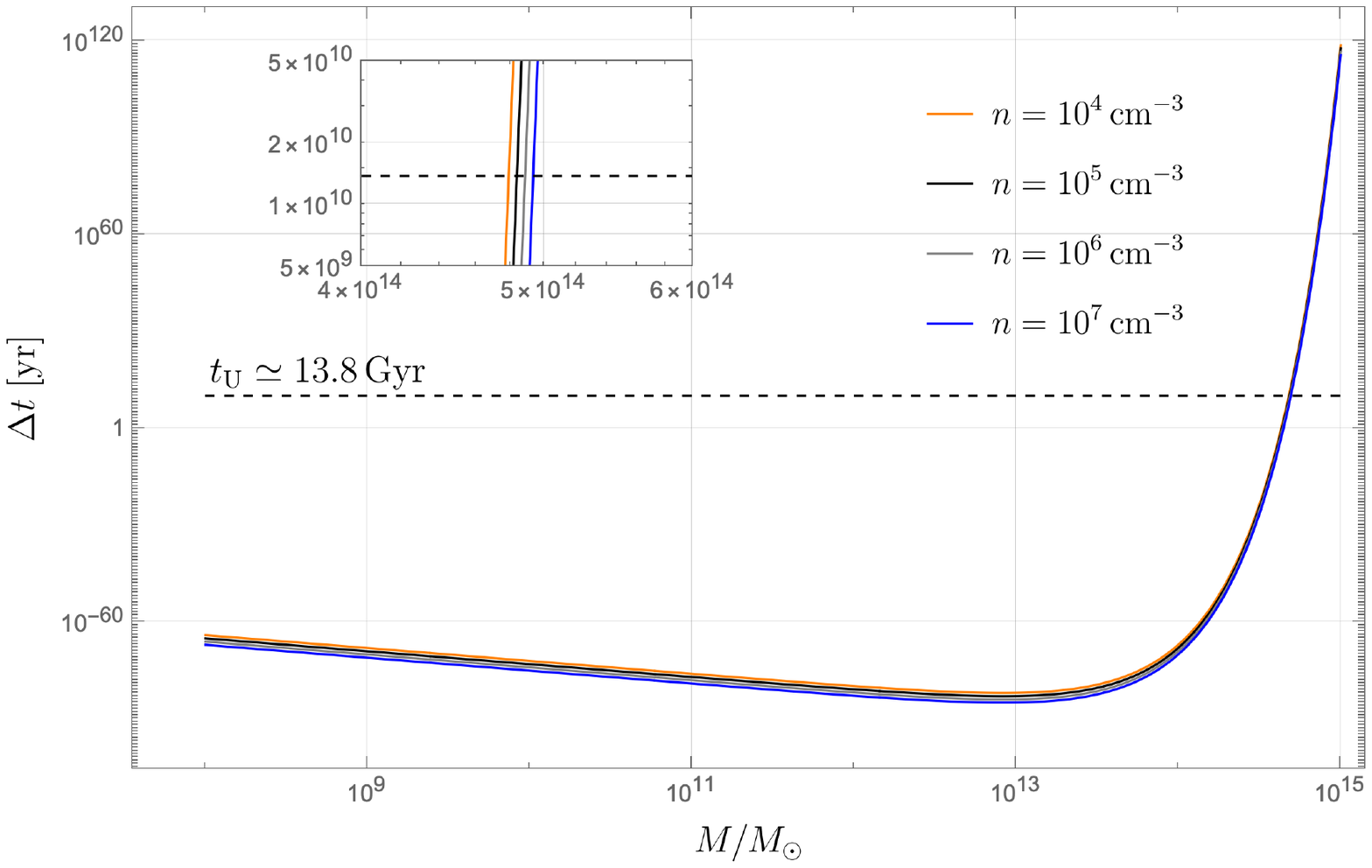}
    \caption{{{\it Discharge timescale due to ionisation of neutral hydrogen}. Plot of the timescale (in years) for hydrogen ionisation discharge of an extremal electrically charged black hole as a function of its mass.  For each mass, $\Delta t$  is obtained from \eqref{eq:Pionintegral}. The different curves correspond to different density assumptions. The dashed horizontal line marks the age of the Universe, $t_U\simeq 13.8\,{\rm Gyr}$.} }
    \label{fig:timescaleionization}
\end{figure}

Putting this together, we {obtain the estimate that} the black hole will discharge unless
\be
\beal
n p \frac{r_h^4}{c} 
= 32 \pi^3 \alpha^7 {\cal N} \frac{L_p^3}{\lambda_e^3} \left ( \frac{Q}{e} \right)^5
\exp\left[ -\frac{8 \pi^2 \alpha^3}{3 } \frac{L_p^2}{\lambda_e^2} \frac{Q}{e} \right ]
&<1\\
\text{i.e.} \qquad 2 \pi^{3/5} \alpha^{7/5} {\cal N}^{1/5} \frac{L_p^{3/5}}{\lambda_e^{3/5}}
\frac{Q}{e}
\exp\left[-\frac{8 \pi^2 \alpha^3}{15} \frac{L_p^2}{\lambda_e^2} \frac{Q}{e} \right]
&< 1\\
\Rightarrow \qquad \frac{Q}{e} = -\frac{15 \lambda_e^2}{8 \pi^2 \alpha^3 L_p^2}
W_{-1} \left ( - \frac{4}{15} \sqrt[5]{\frac{\pi^7 \alpha^8 L_p^7}{{\cal N} \lambda_e^7}}
\right)
& = 4.8 \times 10^{53}
\eeal
\ee
{which gives, for the fiducial neutral hydrogen environment with $n\sim 10^6\, \text{cm}^{-3}$, an order of magnitude lower bound of about  
\begin{equation}
    M\approx 4.5 \times 10^{14} M_\odot
\end{equation}
for avoiding discharge.

Note that the dependence on the assumed hydrogen density enters the argument of the Lambert function as
${\cal N}^{-1/5}$. As before, the Lambert function contains very small 
negative argument and thus the dependence on detailed numerical 
corrections  will be marginal.  
The dominant physical uncertainty is rather the presence of neutral hydrogen atoms, not the exact value of $n$.

Similarly, instead of approximating the interaction 4-volume by a fiducial volume
$r_h^4/c$, if we keep the radial and time dependence of the ionisation rate explicitly, the probability of ionising one hydrogen atom around the black hole in a time $\Delta t$
becomes
\begin{equation}
 \mathcal{P}_{\rm ion} = 4\pi n\,\Delta t
\int r^2 \mathcal{D}(r)\,dr ,
\label{eq:Pionintegral}
\end{equation}
where we have assumed, for simplicity, that the hydrogen density is uniform, and where 
\begin{equation}
    \mathcal{D}(r)=p\, \zeta^2 \,\exp\left(-S_{\rm ion} \left(\zeta^2-1\right)\right)\qquad \text{with }\qquad S_{\rm ion}=\frac{8 \pi^2 \alpha^3}{3 } \frac{L_p^2}{\lambda_e^2} \frac{Q}{e}
\end{equation}
is the probability of ionizing one hydrogen atom at a given distance $r$ from the black hole ($p$ is the expression computed at the horizon and written in Eq. \eqref{ionampl}; while $\zeta=r/r_h$ as previously defined). 
At this point, choosing $\Delta t$ to be the light-crossing time of the black hole, we solve $\mathcal{P}_{\rm ion}=1$ and find that the limiting charge (and mass) of the black hole is 
\begin{equation}
    \frac{Q}{e}\approx 4.7\times 10^{53}, \qquad \text{or} \qquad M\approx 4.4 \times 10^{14} \, M_\odot.
\end{equation}
We note therefore that also in this case, our crude estimate gives the same bound as the one obtained through a proper integral.

At this point, we can also ask the time interval for discharge. This can be answered by solving $\mathcal{P}_{\rm ion}=1$ for $\Delta t$ (refer to Eq.~\eqref{eq:Pionintegral}). The results are shown in figure~\ref{fig:timescaleionization}, and demonstrate that the limiting mass required to discharge one unit within the age of the universe is largely insensitive to the choice of density profile. Furthermore, this limiting mass remains remarkably close to the bound derived using the light-crossing time as the reference scale, showing that even a baseline timescale orders of magnitude smaller than the age of the universe yields a bound of the same order of magnitude. 

}

Given that the most massive SMBH (reported recently in \cite{MMSMBH}) is only 36 billion
solar masses, this bound of nearly half a quadrillion solar masses, 
while comfortably below the total matter in the universe, is nonetheless well beyond 
current conceptual limits for a massive SMBH.

\section{Magnetic charge and extremality}
\label{sec:magnetic}

Since the Reissner--Nordstr\"om solution (from a purely geometric point of view) can be sourced by
the energy momentum tensor for both electric and magnetic charges, we now consider whether 
an extremal magnetic black hole is a realistic prospect in our universe.
Thus far, we have no direct evidence of a magnetic monopole, however, the topology of 
symmetry breaking indicates that monopoles are generically formed.
The existence of a monopole is predicated on a nontrivial $\Pi_2$ of the vacuum, which
in turn is related to a residual unbroken $U(1)$, that we have in electromagnetism. This means that
monopoles are expected to be a by-product of phase transitions in the early universe for 
generic grand-unified symmetry breaking \cite{SimplestGUT-GeorgiGlashow} $-$ 
indeed, the production of monopoles in the early universe was considered so problematic 
it was one of the motivations behind the inflationary paradigm. Since the scale of the mass 
and charge of the monopole are set by the grand unified scale, we expect that discharge 
by pair creation will be enormously suppressed compared to the electric case, and 
indeed the computed rate \cite{Affleck:1981ag} becomes significant only for black holes 
that are already unstable to a process we now discuss.  

The monopole is a nontrivial field configuration in which the field at infinity wraps around the
non-contractable 2-cycles in the vacuum manifold. 
While the details of this field configuration will depend on the details of the gauge group,
the essence of the monopole is captured by the solution of t' Hooft and Polyakov
\cite{tHooft:1974,Polyakov:1974}, where the simplest manifestation of the solution
is constructed. In this model with a scalar and non-abelian vector field, the scalar is in
the adjoint of SO(3), so that there is a very direct and intuitive picture of the sphere
at infinity wrapping the internal sphere of the vacuum. The Lagrangian is:
\be 
\mathcal{L}_\text{m} = \frac{1}{2} D_\mu \mathbf{\Phi}\cdot D^\mu\mathbf{\Phi}
-V(|\mathbf{\Phi}|) -\frac{1}{4 }\mathbf{F}_{\mu\nu}\cdot\mathbf{F}^{\mu\nu},
\ee
where
\be 
\beal
D_\mu \mathbf{\Phi} & =\partial_\mu \mathbf{\Phi}-g  \mathbf{A}_\mu \times \mathbf{\Phi} \\
\mathbf{F}_{\mu\nu} & =\partial_\mu \mathbf{A}_\nu
-\partial_\nu \mathbf{A}_\mu-g \mathbf{A}_\mu \times \mathbf{A}_\nu,
\eeal
\ee
Here, boldface vector notation refers to the internal SU(2) structure, ``$\cdot$'' refers to the
standard inner product on $\mathbb{R}^3$, and the potential 
$V(|\mathbf{\Phi}|)=\lambda(\mathbf{\Phi}\cdot \mathbf{\Phi} -\eta^2)^2/4$ has a minimum 
at $|\mathbf{\Phi}|=\eta$, 
with $\eta$ the symmetry breaking scale. 
The theory contains nonsingular monopoles with magnetic charge $P$ 
(considered to be in $[A \,m]$ units when restoring the natural constants) and 
mass $M_\text{mon}$ given by: 
\be
\label{eq:magneticcharge} 
P=\frac{N}{g},\qquad 
M_\text{mon}\sim \frac{4 \pi \eta }{g },
\ee 
where $N$ is the winding number of the map into the vacuum manifold at infinity. 
For a magnetic charge with winding number $N$, the matter fields, up to a gauge 
transformation, have the asymptotic form 
\be
\mathbf{\Phi}_\infty =\eta\, \hat{\mathbf{e}}(\theta,\phi)\;, \qquad
\mathbf{A}_{_\infty\mu}=(1/g)\,\hat{\mathbf{e}}\times \partial_\mu \hat{\mathbf{e}}
\label{asymptoticform}
\ee
where 
$\hat{\mathbf{e}}=(\sin\theta \cos N\phi,\sin \theta\sin N\phi, \cos\theta)$.
Consequently, the field strength tensor takes the form 
$\mathbf{F}_{\theta\phi}=-\mathbf{F}_{\phi\theta}=(N /g )\sin\theta\,\hat{\mathbf{e}}$, 
which lies entirely in the electromagnetic U(1) subgroup defined by the scalar field 
and precisely reproduces the radial magnetic field expected in the Maxwell–Einstein theory.

The monopole solution, for $N=1$ at least, is obtained by assuming a radial profile
for the magnitude of the scalar and vector fields \cite{Prasad:1975kr}: 
\be
\mathbf{\Phi} = \frac{H(r)}{gr} \hat{\mathbf{e}}(\theta,\phi)\quad , 
\qquad \mathbf{A}_\mu= \frac{(1-K(r))}{gr} \,\hat{\mathbf{e}}\times \partial_\mu \hat{\mathbf{e}}
\label{PS-Ansatz}
\ee
where generically $H$ and $K$ are found numerically, although in the $\lambda \to 0$, or BPS 
\cite{Bogomolny:1975de,Prasad:1975kr} limit, they have an exact analytic form:
\be
H = \frac{r}{r_m} \coth (r/r_m) -1 \quad ; \qquad K = \frac{r}{r_m \sinh(r/r_m)}
\ee
where $r_m$ is a lengthscale representing the effective size of the ``core'' of the 
monopole, where the SU(2) symmetry is restored. At finite $\lambda$, the $1/r$
fall-off in the non-trivial profiles of $\mathbf{\Phi}$ and $\mathbf{A}_\mu$ becomes exponential,
with the width set by $1/\sqrt{\lambda} \eta$ and $1/g \eta$ respectively (for winding number 1).
The monopole therefore has a dense scalar and (nonabelian) vector core with the mass
depending primarily on $g$ and weakly on $\lambda/g^2$ \cite{PGoddard_1978}.
 
Turning to the magnetically charged black hole, if the black hole radius is larger
than this effective core size, the exterior (vacuum) monopole solution can be superimposed 
on the geometry, however, as the black hole mass is dialled down (or the monopole charge 
cranked up) something interesting happens: it is now possible to have part of the monopole 
core peeping out from the horizon $-$ in effect to have the black hole residing within the 
monopole \cite{BHinsideMM}, with the SU(2) fields partially screening the magnetic charge.

Since both magnetically charged Reissner--Nordstr\"om and the dressed black hole
are possible solutions, Lee, Nair and Weinberg tested the stability of the Reissner--Nordstr\"om
solution in the SU(2) theory \cite{Lee:1991}. By considering fluctuations in the scalar and 
vector fields outside the horizon, they showed that the perturbation equations could be
manipulated into a Schr\"odinger form, allowing them to prove the existence of an unstable
mode via a test-function estimation, although they did not compute the actual wave 
function or timescale.
Since we are interested here in timescales, we briefly outline their argument before 
computing the detail of the instability.

First, let us derive an estimate of where the instability might set in for a simple ($N=1$) 
monopole. We might expect this to happen when the
horizon radius drops below the size of the monopole core:
\be
r_h =  M + \sqrt{M^2-P^2} \lesssim r_m \approx \frac{1}{ g \eta }
\label{roughLNWbound}
\ee
however, assuming the charge is entirely magnetic and due to the monopole, we
have $P = 1/g$, hence we deduce that an instability can only emerge if 
$1/g\leq M<{\cal O} (1/g\eta)$. Rearranging \eqref{roughLNWbound} we therefore see that
an estimate of the emergence of an instability will be when the black hole has a mass
approximately below
\be
M \lesssim \frac1{2 g \eta} + \frac{\eta}{2g}
\label{roughMbound}
\ee

To test this intuition, Lee, Nair, and Weinberg, motivated by the Prasad--Sommerfeld 
Ansatz \cite{Prasad:1975kr}, considered spherically symmetric perturbations around 
the monopole vacuum \eqref{asymptoticform} of the form \eqref{PS-Ansatz} but with 
$H$ and $K$ now functions of $r$ and $t$. They found that the instability resided in
the vector component:
\be
K(r,t) \sim e^{\Omega t} u(r)
\ee
with an equation of motion
\be
-\Omega^2 u = -\frac{d^2 u}{d r^2_\star} + U(r_\star) u
\label{LNWperturbation}
\ee
where $r_\star = \int dr/f(r)$ is the tortoise coordinate ($f(r) = 1 - 2M/r + P^2/r^2$ being
the Reissner Nordstr\"om potential) and 
\be
U = -\left ( 1 - \frac{2M}{r} + \frac{P^2}{r^2} \right) \frac{(1 - g^2 \eta^2 r^2)}{r^2}
\ee
with $r=r(r_\star)$ implicitly understood.

With this perturbation equation in Schr\"odinger form, it is easy to see the instability corresponds
to a bound state solution, which Lee et al.\ argued the existence of 
using trial eigenfunctions.
Since $\eta \sim 10^{-4}$ for a typical grand unified theory (GUT) breaking scale, we see that $U$ is negative for $r_h\lesssim1/g\eta$ in agreement with the 
heuristic argument above.

When $N>1$, the estimate above, \eqref{roughMbound}, for the onset of the instability 
becomes
\be
M \lesssim \frac{\sqrt{N}}{2\eta g} + \frac{2 \pi \eta N \sqrt{N}}{g}
\label{eq:massconstraints}
\ee
which also implies a bound of the winding number of $N\lesssim 1/\eta^2$. In detail, the 
potential in the instability equation is modified by replacing the $g^2 \eta^2 r^2$ term in brackets 
with $Ng^2 \eta^2 r^2$ \cite{Lee:1991}. Strictly, \eqref{LNWperturbation} only picks up an unstable
s-mode, however Lee et al.\ put forward a heuristic picture in which, as the fields condense
they localize at points on the horizon, and they conjectured that these regions could grow into distinct 
lumps and detach as individual magnetic monopoles, thus reducing the total magnetic charge (a 
related process was considered by Maldacena \cite{Maldacena:2020skw}). 
Eventually, when only a single unit of charge remains, the horizon may shrink further by 
Hawking evaporation, exposing the monopole core. 

To see how the instability depends on the physical parameters, consider the rescaled 
equation
\be
\frac{1}{\hat{f}}\frac{d~}{d\hat{r}}\,\frac{1}{\hat{f}}\frac{d~}{d\hat{r}}\, u(\hat{r})  = \left [\hat{\Omega}^2 + 
\frac{(1 - \hat{ r}^2)}{{\hat{ r}}^2} \right] u(\hat{r}) = \hat{U}(\hat{r}) \,u(\hat{r})
\ee
where
\be
\hat{r} = g \eta r \;\;, \quad \hat{\Omega} = \Omega /g \eta \;\;, \quad
\hat{M} = g \eta M \;\;, \quad
\hat{f} = \left ( 1 - \frac{2\hat{M}}{\hat{ r}} + \frac{N^2 \eta^2}{\hat{ r}^2} \right) 
\ee
Clearly, for $\eta\sim 10^{-4}$, the lower bound on $\hat{M}$, $N\eta$, is small, and expanding
in $\hat{M}$ shows that the minimum of $\hat{U}$ occurs at $r_\text{min} \in [2{\hat M}, 3\hat{M})$,
(with the lower bound of $r_{\rm min}$ occurring at the extremal limit, and the upper at close to zero charge), 
with a minimum value of 
\be
\hat{U}_{\text{min}} \in \left [-\frac{1}{16 \hat{M}^2}, -\frac{1}{27 \hat{M}^2}\right)
\ee
(again with the lower bound corresponding to the extremal limit). Trial eigenfunctions
show that the eigenvalue corresponding to the instability is well approximated by
$\hat{\Omega} = a \sqrt{|\hat{U}_{\text{min}}|}$, where $a$ is very close to $1$.
Unravelling the rescaling, we therefore expect that the timescale of the instability
$\Omega \propto 1/M$, and independent of $N$ away from the extremal limit. 
At, or close to, the extremal limit, since $M$ is bounded below by the charge, we
expect that $\Omega$ will be roughly inversely proportional to $N$ through the dependence
of $M$. We therefore do not need to perform a large exploration of parameter space, as 
the behaviour of the rescaled parameters informs us of how the eigenvalues
depend on the black hole parameters.

\begin{figure}[ht]
\centering
\includegraphics[width=0.8\linewidth]{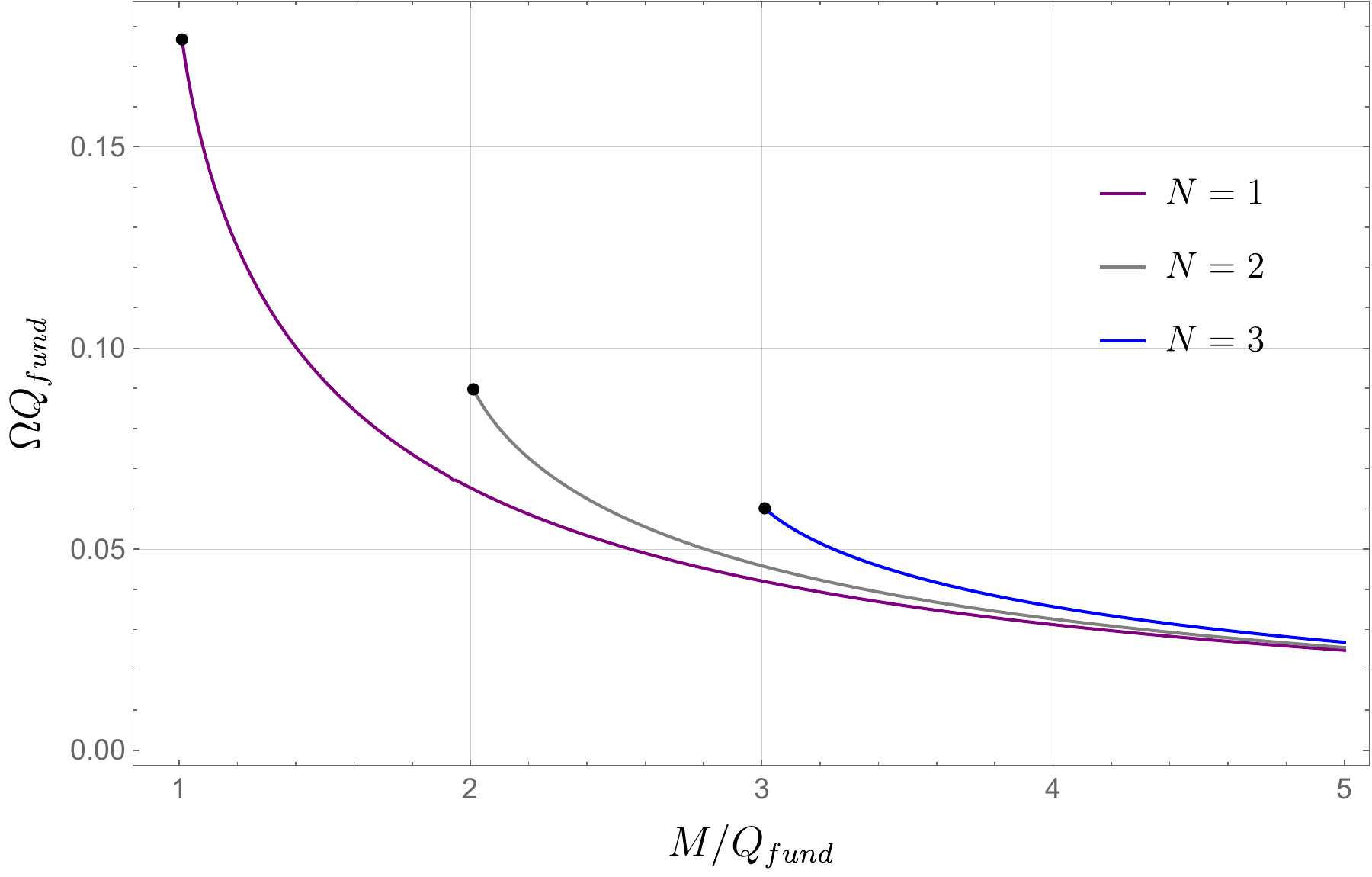}
\caption{\textit{Instability of magnetically charged black holes.} Plot of $\Omega$ for a 
magnetically charged black hole as a function of its mass $M$, expressed in units of the 
fundamental magnetic charge. For reasons of dynamic range we only display results for 
low $N$. Each curve corresponds to a different winding number $N$, 
as indicated in the legend. The black dots mark the extremal limit: the minimum 
mass possible for a black hole with the given charge. Accordingly, the curves do not 
extend beyond this point. We adopt natural units with $G=c=\hbar=4\pi\epsilon_0=1$, 
and set $\eta=10^{-4}$ and $g=1$, so that the fundamental magnetic charge is unity 
and the charge of the black hole is $Q=N$. The mass of the extremal black hole equals $N$.} 
\label{fig:FreqInst-Magnetic}
\end{figure}

We solved the wave equation for $u$ using a shooting method, expanding the solution
at both boundaries in terms of $\Omega$, integrating in to a matching intermediate
radius, where demanding that the wave function is $C_1$ determines $\Omega$. {For more details about the numerical implementation we refer the reader to Appendix \ref{app:a}}.
Figure~\ref{fig:FreqInst-Magnetic} shows the instability frequencies obtained
by solving the perturbation equation \eqref{LNWperturbation} for a range of
black hole masses and winding numbers $N$. 

The plot in figure~\ref{fig:FreqInst-Magnetic} shows clearly that at fixed mass, the higher charged black hole is more unstable, as
we might expect. Clearly, at fixed charge, the extremal black hole is the most unstable, as
expected, as this is the smallest horizon size. As the mass increases, the black hole becomes
less unstable in line with expectations: once the condition in Eq.~(\ref{eq:massconstraints}) 
is no longer satisfied, the black hole becomes stable and $\Omega$ drops to zero.
Extremal black holes also become more stable as the charge increases, until the maximum 
charge is attained. Since extremal black holes are not expected to evaporate (having
zero temperature) we see that above this critical threshold, the extremal magnetic black
hole is stable.

Note that the instability condition on $M$, \eqref{eq:massconstraints}, for the extremal
black hole can be rewritten as 
\be 
\frac{2 \sqrt{N}}{g\eta} \left [ \left(x-\frac{1}{4}\right)^2 
+ \left ( \frac\pi4- \frac{1}{16} \right ) \right ]\ge 0,
\ee 
with $x^2=N \pi^2 \eta^2 $. This constraint is always satisfied, hence the critical threshold
for the mass of a stable extremal black hole is determined by the upper limit on the charge
multiplicity: $N\lesssim 1/\eta^2\approx 10^6-10^8$ for typical GUT scenarios.
Therefore, the existence of stable, magnetically charged extremal black holes in our universe
boils down to the viability of a black hole containing $N\gtrsim 10^6 $ monopoles. The 
immediate problem with this is that we expect a period of inflation to dilute the monopole
density to at most ${\cal O}(1)$ monopole per Hubble volume. Therefore, any process producing such
a highly charged black hole would require a black hole to be present at or before inflation, 
which would most likely require the black hole to be formed via a tunnelling process. 

Black hole pair creation was considered some time ago in the context of decay of cosmic 
strings \cite{Hawking:1995zn,Emparan:1995je,Eardley:1995au,Gregory:1995hd}
or magnetic fields \cite{1995Euclideanactions}
where the C-metric was used to construct a Euclidean instanton interpolating between
the initial string or magnetic field to a pair of black holes accelerating apart.
This type of nucleation would require some specific background that, while possible if 
cosmic strings are formed during the phase transition, will not produce black holes
of sufficient size to accrete such a high monopole charge.
The relevant process to consider is where a black hole spontaneously appears in the
universe, as considered by Bousso and Hawking in \cite{1996PairBHs}
{via a Hawking-Moss \cite{Hawking:1981fz} process}. Such a process
reduces entropy, hence is strongly Boltzmann suppressed; for nucleation of an
uncharged Nariai black hole -- the type considered by Bousso and Hawking --
this amplitude is:
\be{\cal P}_{BH} \sim e^{-\pi\ell^2/3G},
\ee
where $\ell = \sqrt{\Lambda/3}$ is the de Sitter lengthscale of inflation.
Since $\ell \sim 10^5$ Planck lengths, the nucleation rate of uncharged 
Nariai black holes is extremely low.

{The reason Bousso and Hawking computed the Boltzmann factor for the Nariai black hole 
was that the general Schwarzschild de Sitter (SdS) solution has a mismatch
of temperature between the black hole and cosmological horizons, meaning that
the ``natural'' periodicity of Euclidean time does not match at each horizon, thus a conical 
singularity is present at one or both of the horizons in the Euclidean section.} 
However, the presence of a conical 
singularity does not preclude a finite Euclidean action \cite{Banados:1993qp,Gregory:2013hja},
and finite action singular geometries contribute to tunnelling processes such as decay
of the false vacuum \cite{Burda:2015isa,Burda:2015yfa}.

{ Instead, here we consider a completely general process of 
spontaneous creation of a black hole, using the approach described in
\cite{Gregory:2020hia} where completely general 
Hawking--Moss instantons tunnelling from a lower to higher cosmological constant
were considered with both seed and remnant black holes. Since we are tunnelling
from pure de Sitter to SdS at the same cosmological constant,
we do not change the background $\Lambda$, rather, we consider the jump from 
pure de Sitter to an SdS solution of arbitrary mass.
The semi-classical amplitude exponent is defined as the discrepancy between the action of
the nucleated SdS solution and the initial inflationary background,
assumed to be pure de Sitter in \cite{1996PairBHs}.
We now briefly outline this computation, including charge for the black
hole so as to also compare with the instantons considered by Bousso and 
Hawking. 

The Euclidean action of the de Sitter charged black hole is given by\footnote{{Note, for brevity 
we will not include the Gibbons-Hawking-York boundary term as we will integrate over 
the conical deficits. A more careful treatment can be found in \cite{Gregory:2013hja}}}
\be
I_E = \frac{-1}{16\pi} \int \left (R - 2\Lambda - F_{ab}^2\right ) \sqrt{g} d^4 x
\ee
and consists of three main pieces: the bulk integral between the two horizons, and the two
horizon integrals of the delta-function in the curvature if a conical defect is present.
As described in \cite{Gregory:2013hja}, the integral $\int \sqrt{g}Rd^4x $ over a conical defect
is finite and proportional to the deficit angle and area of the orthogonal space --
the horizon:
\be
\lim_{\epsilon\to0} \int_{r -r_i \leq \epsilon} \sqrt{g} R d^4 x
= 2 \pi \delta_i {\cal A}_i,
\ee
where the index $i$ labels the horizon, ${\cal A}_i$ is the area of that horizon, and 
$2\pi \delta_i$ is the deficit angle at the horizon. In terms of the ``natural'' periodicity
$\beta_i = 4\pi / |f'(r_i)|$ of the horizon and the actual periodicity of Euclidean time, the conical deficit is
\be
2 \pi \delta_i = \frac{2\pi}{\beta_i} (\beta_i - \beta).
\ee
Putting together these contributions with the bulk integral, one obtains
for the geometry
\be
\frac{-1}{16\pi} \int \left ( R - 2\Lambda \right ) \sqrt{g} d^4 x
= - \frac{\beta}{2 \ell^2} (r_c^3-r_h^3)
- \frac{1}{4} ( {\cal A}_h + {\cal A}_c ) + \frac{\beta}{4} 
\left ( \frac{{\cal A}_h}{\beta_h} + \frac{{\cal A}_c}{\beta_c}
\right )
\ee
together with the electromagnetic contribution
\be
\frac1{16\pi} \int F^2 \sqrt{g} d^4 x = - \frac{P^2\beta}{2} \left ( \frac1{r_c} - \frac1{r_h} \right )
\ee

Now one substitutes in the values of the natural periodicities and areas:
\be
\beal
\frac{{\cal A}_i}{\beta_i } &= r_i^2 |f'(r_i)| = 2 \left | M -  \frac{P^2}{r_i} -  \frac{r_i^3}{\ell^2} \right | \\
\Rightarrow \qquad \left ( \frac{{\cal A}_h}{\beta_h} + \frac{{\cal A}_c}{\beta_c} \right ) &=
2 \left ( M -  \frac{P^2}{r_h} -  \frac{r_h^3}{\ell^2} \right ) 
- 2 \left ( M -  \frac{P^2}{r_c} -  \frac{r_c^3}{\ell^2} \right )\\
&= 2 P^2 \left ( \frac{1}{r_c} - \frac{1}{r_h} \right ) + \frac{2}{\ell^2} \left ( r_c^3 - r_h^3 \right )
\eeal
\ee
to obtain:
\be
\beal
I_E & = - \frac{\beta}{2 \ell^2} (r_c^3-r_h^3)
- \frac{1}{4} ( {\cal A}_h + {\cal A}_c ) + \frac{\beta}{4} 
\left ( \frac{{\cal A}_h}{\beta_h} + \frac{{\cal A}_c}{\beta_c}
\right )
- \frac{P^2\beta}{2} \left ( \frac1{r_c} - \frac1{r_h} \right ) \\
& = - \frac{1}{4} ( {\cal A}_h + {\cal A}_c ) ,
\eeal
\label{chargedinstantonaction}
\ee
i.e.\ the serendipitous cancellation of the arbitrary time-periodicity discovered for the uncharged
SdS black hole in \cite{Gregory:2013hja} persists even for the charged black hole!
We now apply this result to uncharged and charged
instantons in turn.

For uncharged black holes, the roots of the cubic lapse function can be 
neatly written in terms of an angle encoding the mass: $\cos(3b) 
= 3 \sqrt{3} M/\ell$, with $b=0$ being the Nariai limit 
and $\pi/6$ being zero mass. In terms of this angle,
\be
r_h = \frac{2\ell}{\sqrt{3}} \cos\left ( \frac\pi3 + b \right ) \quad
r_c = \frac{2\ell}{\sqrt{3}} \cos\left ( \frac\pi3 - b \right ) \;\;\; \Rightarrow \;\;\;
I_E = \frac{2\pi \ell^2}{3} \left [ \cos(2b) - 2 \right ]
\ee
hence the exponent of the semi-classical amplitude, $e^{-{\cal B}}$, is 
\be
{\cal B} =I_E - I_{dS} = \frac{\pi \ell^2}{3}\left(2\cos( 2b ) -1 \right),
\label{eq:unchargedB}
\ee
which continuously interpolates from zero (no tunnelling, no black hole) to 
the Bousso-Hawking result ${\cal B}_{BH}=\pi\ell^2/3$. 
Clearly, the nucleation of a small black hole is preferred over a larger one, however the black hole should have
sufficient volume to absorb the high number of monopoles if we wish to create a stable extremal black hole.

Turning to charged instantons, we follow the same procedure as
Bousso and Hawking: spontaneously nucleating a charged black hole 
from pure de Sitter space. As before, Bousso and Hawking required the 
periodicities of the two horizons to be identical, which constraint demands that
$P=M$. Note, once there is a cosmological constant, $M=P$ is no longer
the extremal limit, instead there are two critical relations for $P$ and $M$
corresponding to the extremal ($M_{ex}$) and Nariai ($M_{N}$) limits:
\be
M_{ex}^2 = 1 + 36 P^2 - \frac{(1 - 12 P^2)^{3/2}}{3 \sqrt{6}},\qquad
M_N^2 = 1 + 36 P^2 + \frac{(1 - 12 P^2)^{3/2}}{3 \sqrt{6}}
\ee
Since $M_{ex}<P$ the Bousso-Hawking limit is close to, but still subextremal,
and was dubbed a  {\it lukewarm} instanton.

The exponent in the amplitude, as before, is the difference in 
cumulative horizon areas, however, with the horizon radii being solutions
to a quartic, the expressions are not particularly illuminating, so
instead we plot the comparison of the general periodicity action to the
lukewarm Bousso-Hawking instanton in figure~\ref{fig:nucleation-charged}.
What this figure shows is that in stark contrast to the uncharged case,
the Bousso-Hawking nonsingular instanton is now the minimal action one.
Further, the lower the charge, the more strongly localised this minimum is.
We should however mention one concern about this computation, namely 
the background subtraction. As in \cite{1996PairBHs} we subtract the pure de Sitter 
background, however, this is topologically trivial from the perspective
of the field theory in which the monopoles live. While we might expect
that the lukewarm instanton should be used if we are considering 
spontaneous nucleation of a black hole around a distribution of monopoles,
perhaps one ought to have the background also include these 
monopoles. 
\begin{figure}[htbp]
    \centering
    \includegraphics[width=0.9\linewidth]{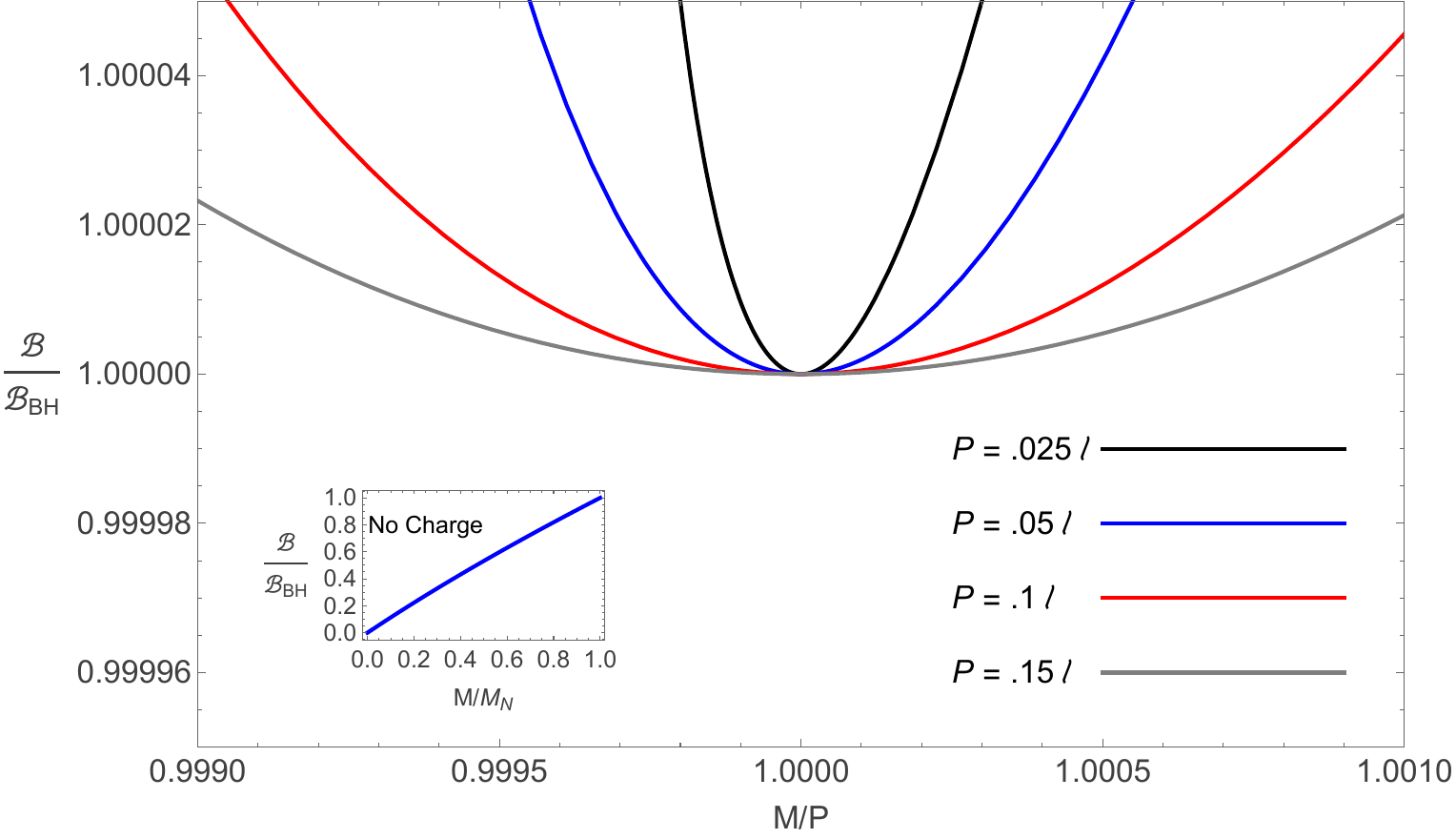}
    \caption{{{\it Charged instantons.} A comparison of the charged instanton exponent to the 
    Bousso-Hawking lukewarm instanton. We see that the Bousso-Hawking
    lukewarm instanton always has the lowest action, and the minimum
    becomes more marked at lower charge. Inset is a plot of the comparison for
    zero charge, where the general (non-Bousso-Hawking) instanton always has lower action.}}
    \label{fig:nucleation-charged}
\end{figure}

}

{Now consider the probability of a black hole nucleating with sufficient 
volume that it is created at the high charge required for stability. For this,
we first estimate the monopole density, then infer the required volume and
compare both instanton amplitude results.
Defects, such as the monopole, are
formed during a phase transition via the Kibble mechanism 
\cite{Kibble_1976,Preskill:1979zi,Kibble:1990rg}, where it is assumed
that the order parameter (determined by the underlying field) is uncorrelated 
over distances larger than some scale $\xi$, the correlation length. 
These uncorrelated regions will then typically have nontrivial topological
winding numbers at larger scales, that leads to the formation of 
topological defects, such as monopoles. The monopole density is then
approximately $n \sim a \,\xi^{-3}$, where $a$ is a factor related to
the expectation that the uncorrelated regions will conspire to give a
topological winding, and is expected to be of order $1-10^{-1}$. 
For a GUT model, the scale of fluctuations in $\rm \Phi$ is 
given by the inverse of the scalar mass $\xi\sim 1/\sqrt{2\lambda}\eta$,
where $\eta$ is the symmetry breaking scale and $\lambda$ a self-coupling. 
Assuming a GUT correlation length of $\xi \sim 10^{3}$ (in Planck units)
an order-of-magnitude estimate for the initial monopole number density 
is $10^{-9}-10^{-10}$ per Planck volume. Thus, if a spontaneously nucleated
black hole is to capture $\sim 10^6$ monopoles, then we require its 
volume to be ${\cal V} \sim 10^6 / n_m = 10^{15}-10^{16}$ Planck volumes. 
The displaced volume of the black hole is ${\cal V} = 4\pi r_h^3/3$, leading to
$r_h \sim 5 \times 10^4 L_p$. Assuming inflation takes place close to this
grand unified scale, then the Hubble scale is $\ell \sim 10^6-10^8$, hence the
radius is around, or a bit under, $1\%$ of the Hubble scale. 
This estimate assumes the monopole density is uniform, 
hence neglects possible inhomogeneities or clustering in the monopole distribution.

For small mass, both the uncharged amplitude \eqref{eq:unchargedB} and the Bousso-Hawking
exponents have the same behaviour: ${\cal B} \sim 2 \pi M \ell\approx \pi r_h \ell$. 
This exponent is at best of order $10^{10}$, hence the amplitude for black hole nucleation
at this size and charge is negligible, rendering the abundance 
of black holes nucleated via tunnelling negligible even when more general channels are included. 
  }

Finally, it is worth noting that the evidence for GUT's is indirect, and there is no experimental
evidence for monopoles. However, upcoming experiments such as Hyper-Kamiokande \cite{hyperKamiokande} 
and DUNE \cite{Dune} will significantly enhance sensitivity to proton decay, a key prediction of GUTs. 
Their results may soon provide decisive evidence for or against GUT-scale physics, with direct 
implications for the existence of magnetic monopoles (which are also the target of direct searches 
by the MoEDAL experiment \cite{moedal}) and the associated black hole solutions.

\section{Discussion}
\label{sec:conclusions}

We have shown that charged extremal astrophysical black holes are inherently 
unstable once interactions with the environment and cosmological history are taken into account.
For electric black holes, Schwinger pair production in the vicinity of a charged black hole 
must be weighted by computing not only the rate for a single electron-positron pair, but the 
integrated production rate within the spatial region where the electric field remains 
approximately of the same order of magnitude of its value at the event horizon.
We considered a representative spatial volume for discharge set by the black hole
radius, and explored the relation between the mass of the black hole and discharge time.
Once vacuum polarization becomes inhibited, the black hole can also discharge via ionisation of neutral
atoms in its vicinity. We focussed on Hydrogen to get a further bound on the likely discharge of
an extremal black hole, which pushed up the lower bound on mass to ${\cal O} (10^{14} M_\odot)$.

For extremal magnetic black holes that will not discharge via pair production, we
computed the characteristic timescales associated with the Lee--Nair--Weinberg instability
(an instability that arises when the monopole fields begin to leak out through the black 
hole horizon). We found that stable extremal magnetic black holes must have a charge of 
$10^6-10^8$ magnetic monopoles. Black holes with such large magnetic charges could only 
plausibly form before, or at the start of, inflation. 
{We revisited the Bousso-Hawking
amplitude for pair creation, including the possibility of singular geometries with
finite action, and demonstrated that while the probability of spontaneously nucleating
uncharged black holes is strongly enhanced over the Bousso-Hawking rate, the charged
instantons are dominated by the Bousso-Hawking lukewarm instanton, with both results converging 
at low $M$.
The requirement of
large magnetic charge places a constraint on the mass of the nucleated black hole that
renders the probability of formation still vanishingly low.}

One might ask what happens in the case of dyonic black holes. If electric discharge is so
efficient, but magnetic discharge suppressed $-$ could a black hole with sufficient magnetic
charge somehow stabilise? For an extremal dyonic black hole, the mass $M$ is related to the 
magnetic $P$ and electric $Q$ charges via $M^2= Q^2+ P^2$. 
The resulting electric field at the horizon is given by 
\be 
\begin{aligned} 
|{\bf E}|=&\frac{c^4}{\sqrt{4\pi\epsilon_0 G^3}M^2}\sqrt{\left(4 \times 10^{60}\,[kg]\right) 
\frac{M^2}{M_\odot^2}- \frac{\pi \hslash^2}{\mu_0  e^2  G}\, N^2}\\
&\simeq \frac{c^4}{\sqrt{4\pi\epsilon_0 G^3}M^2}\sqrt{\left(4 \times 10^{60}\,[kg]\right) 
\frac{M^2}{M_\odot^2}- \left(2 \times 10^{-14}\,[kg] \right)\, N^2}.
\end{aligned}
\ee 
We observe that, due to the relative strength of the two terms under the square root, 
the inclusion of a magnetic charge does not significantly alter the conclusions drawn in 
section~\ref{sec:electric}: the electric field remains strong enough in the supermassive 
regime to ionise hydrogen, and thus discharge still occurs. 

It is natural to ask whether these conclusions are modified with the inclusion of 
a dark sector \cite{Essig:2013lka}. For black holes carrying charge under a hidden, unbroken $U(1)$, we can 
reframe our results in terms of a more general charge:mass ratio $z$ of the lightest 
elementary dark particle. We expect our results will be broadly similar for large $z$;
from \eqref{exp-prefactor} and \eqref{fullexpansion}, we get a limit on the charge
of the black hole in terms of the charge, $q$ and Compton wavelength, $\lambda$ of
the lightest elementary dark particle:
\be
\frac{Q}{q} \approx \frac{\lambda^2}{\pi^3 L_p^2} \log z.
\ee
From this, we expect that it is only for relatively low charge:mass rations, $z\lesssim {\cal O} (10^2)$ that
our analysis will require significant correction.
This means that our conclusions should be valid for a wide range of charge:mass ratio of the dark 
sector and so dark sectors containing sufficiently light and/or strongly coupled charged fermions 
are expected to discharge efficiently, in close analogy with the ordinary electromagnetic case.
If instead the lightest dark charged fermion is heavy or very weakly coupled, so that 
$z \lesssim 1$, Schwinger discharge is very suppressed, following the results of 
Lin and Shiu \cite{Lin:2024jug}. This allows (near)extremal dark charged black holes 
to persist for much longer. This possibility has been explored recently in the context 
of primordial black holes \cite{2025PhRvD.112l3529S} and shows that hidden charges represent a
model-dependent loophole to the Standard Model discharge bounds, thanks to their different charge:mass ratios.

We have focussed on charged black holes since they are often the most considered 
in high-energy and quantum gravity studies. However, rotating black holes
are distinct from the charged case as astrophysical black holes are observed 
to rotate very rapidly \cite{LIGOcatalog,ExperimentalSpin}. 
Limits on spinning up a black hole via accretion disks have been studied
\cite{1974Thorne,Mummery:2025zak}, as well as angular momentum loss via the superradiance effect \cite{Zel'dovich,superradiance-review,1973Starobinski,1974JETPStarob-Churilov,Baybay:2025kvb}. Additionally,
long lived boson clouds from ultralight dark matter\cite{ultralightDM} can 
also draw angular momentum from the black hole 
\cite{Brito:2014wla,East:2017ovw,superradiance-review}.
We also suggest a further source of scattering that to the 
best of our knowledge has not yet been explored in the literature: the interaction of rotating 
black holes with the stochastic gravitational wave background (for which there has recently been 
evidence via pulsar timing arrays in the nanohertz band \cite{nanograv}). 
Indeed, some frequencies of this background overlap with the superradiant window of 
any extremal Kerr black hole, since the angular velocity of an extremal Kerr horizon is 
\be
\Omega_h=\frac{ c^3}{2GM}\sim 10^5 \frac{M_\odot}{M }\,[\text{Hz}],
\ee 
leading to a net, yet presumably very low, spin down of black holes.

In conclusion, the existence and possible observation of electrically or magnetically 
charged extremal black holes in our universe appears highly implausible. 
In contrast, while exact extremality may be ruled out, near-extremal rotating black holes 
remain viable astrophysical objects. The question remains whether they can be close enough to 
extremality to offer a promising setting for exploring quantum gravitational effects.

\noindent{\bf Acknowledgments}

We would like to thank Sebastian Ellis for helpful conversations.
CC is supported by King’s College London through an NMES funded studentship. 
RG is supported in part by STFC (ST/X000753/1) and in part by the Perimeter 
Institute for Theoretical Physics.
Research at Perimeter Institute is supported in part by the Government of Canada
through the Department of Innovation, Science and Economic Development and by
the Province of Ontario through the Ministry of Colleges and Universities.
This work was also performed in part at the Aspen Center for Physics, which is 
supported by National Science Foundation grant PHY-2210452 (RG).

\appendix

\section{Numerical computation}
\label{app:a}
To determine the numerical results presented in section~\ref{sec:magnetic}, we first perform a near horizon expansion of the function $u(r)$ in Eq.~\eqref{LNWperturbation}. At leading order, we assume a scaling behaviour of the form $(r-r_h)^\beta$. Substituting this ansatz into the field equation yields two roots for $\beta$; we select the value corresponding to ingoing waves to satisfy the physical boundary condition at the black hole horizon:
\begin{equation}
    \beta = - \frac{i \Omega\, (M+\sqrt{M^2-Q^2})^3}{2 \sqrt{(M^2-Q^2) (-Q^2+2 M (M+\sqrt{M^2-Q^2}))}}.
\end{equation}
We then extend this to a higher order power series expansion near the horizon, parametrised as:
\begin{equation}
    u(r) = (r-r_h)^\beta \sum_{i=0}^\infty c_i (r-r_h)^i,
\end{equation}
where the coefficients $c_i$ are determined by solving the field equation order by order close to $r=r_h$. These coefficients depend on $\Omega$ and are expressed in terms of the first coefficient $c_0$, which remains a free normalization constant.

We implement a parallel procedure asymptotically at infinity using a Frobenius expansion:
\begin{equation}
    u(r) = e^{k r} r^\gamma \sum_{j=0}^\infty \frac{d_j}{r^j}.
\end{equation}
Solving the leading order equation at infinity yields two possible values for $k$. To satisfy physical boundary conditions, we choose the root that enforces outgoing waves:
\begin{equation}
    k = -\sqrt{N g^2\eta^2-\Omega^2}.
\end{equation}
Solving the equation at the next-to-leading dominant order determines the exponent $\gamma$:
\begin{equation}
    \gamma = \frac{-g^2 MN \eta^2+2 M \Omega^2}{\sqrt{g^2 N \eta^2-\Omega^2}},
\end{equation}
while subsequent subleading orders yield the expansions of the coefficients $d_j$ in terms of $d_0$ and $\Omega$.

Finally, we compute the frequencies $\Omega$ using a shooting method. Taking the series expansions at the horizon (evaluated at $r_h + \epsilon$) and at infinity as initial data, we integrate the differential equations numerically from both boundaries toward an intermediate matching radius $r_\text{cut}$. This routine provides two numerical evaluations of $u(r)$ at $r_\text{cut}$: one parametrised by $(\Omega, c_0)$ from the outward horizon integration, $u_h$, and another parametrised by $(\Omega, d_0)$ from the inward asymptotic integration, $u_\infty$. 

Because the absolute normalizations of $c_0$ and $d_0$ are arbitrary, we can enforce the simultaneous continuity of the function and its first derivative by requiring the vanishing of their Wronskian at $r_\text{cut}$:
\begin{equation}
    W(u_\infty, u_h) = u_\infty \frac{du_h}{dr} - u_h \frac{du_\infty}{dr} = 0.
\end{equation}
By searching for roots where both the real and imaginary parts of the Wronskian vanish, we uniquely determine $\Omega$. The real part of $\Omega$ obtained via this method is consistently compatible with zero at machine precision ($\sim 10^{-26}$). Furthermore, evaluating the residual value of the Wronskian allows us to monitor our numerical error, which remains bounded on the order of $10^{-14}$.

The exact numerical values obtained depend on several algorithmic parameters: the numerical position of infinity, the matching radius $r_\text{cut}$, the initial horizon offset $\epsilon$, and the truncation orders of the boundary series expansions. To confirm the stability and robustness of our eigenvalues, we systematically varied each of these parameters. 

\begin{figure}[htbp]
    \centering
    \includegraphics[width=0.8\linewidth]{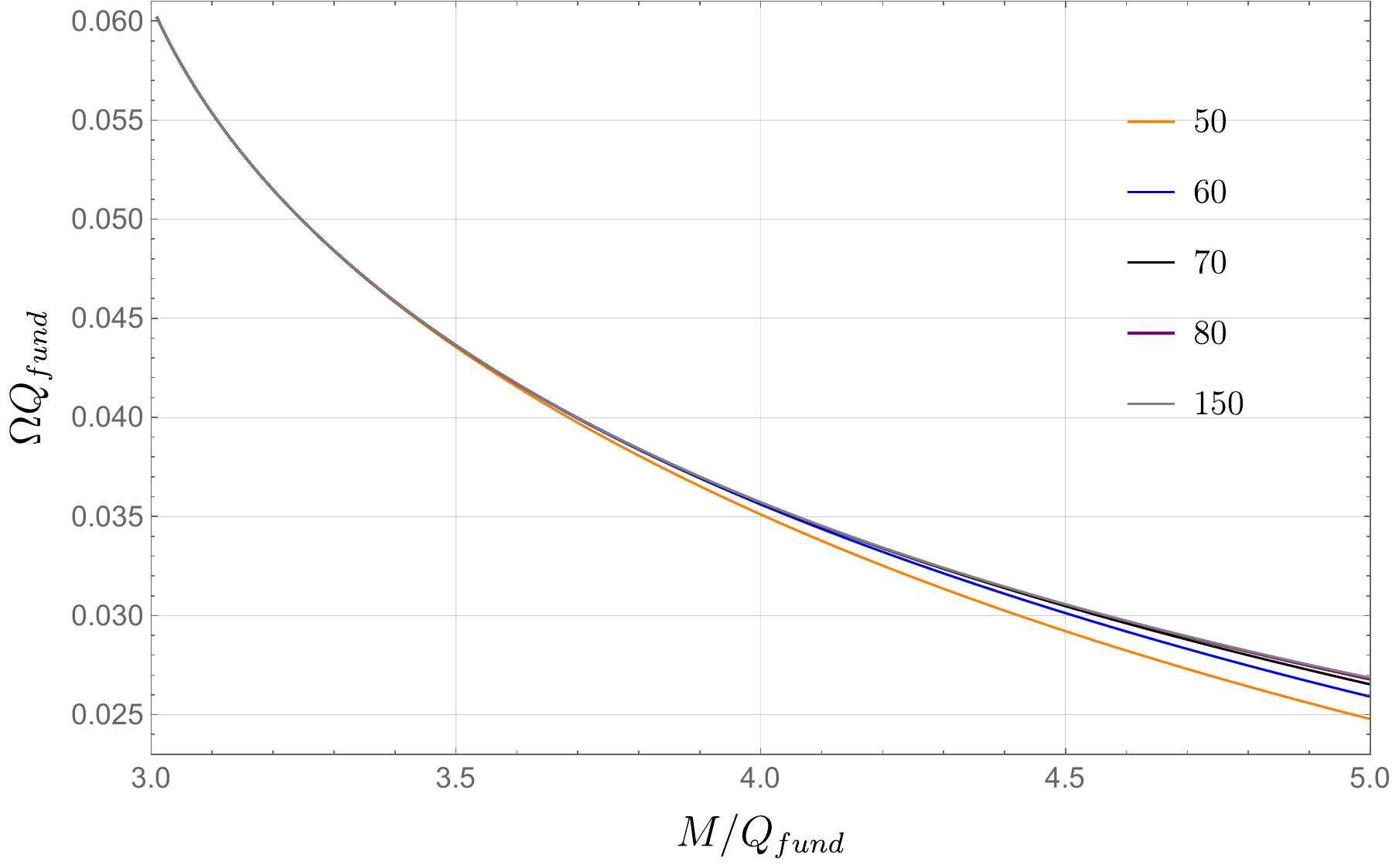}
    \caption{{{\it Numerical convergence}. Behaviour and convergence of the numerical solution for $\Omega$ as a function of the black hole mass. The distinct lines represent results obtained for different numerical approximations of infinity, evaluated here for $N=3$.}}
    \label{fig:stabilityInf}
\end{figure}

First, we verified that shifting the matching point $r_\text{cut}$ does not alter the final outcome. Second, we examined the dependence on the parameter $\epsilon$, which defines the initial displacement from the horizon where numerical integration begins ($r_{\text{start}} = r_h + \epsilon$). Holding the near horizon expansion fixed to order $\mathcal{O}(r-r_h)^8$, we found that decreasing $\epsilon$ does not affect the eigenvalues, demonstrating excellent numerical stability near the horizon. Third, we investigated the convergence of our results with respect to the chosen numerical value for infinity, using the asymptotic expansion up to corrections of order $\mathcal{O}(1/r^8)$. The numerical cutoff for infinity was found to be the parameter to which the system was most sensitive. To illustrate this behaviour, figure~\ref{fig:stabilityInf} displays the convergence of the solution for the $N=3$ case as the numerical value of infinity is systematically varied.

An identical stability analysis was performed for all other charged black hole configurations considered in this work to guarantee the convergence of our data. By optimizing both the cutoff boundaries and the truncation orders of the asymptotic expansions, we ensured strict convergence across the entire parameter space.

\end{document}